\documentclass[12pt]{article}
\usepackage{amsmath}
\usepackage[dvips]{graphicx}
\usepackage{epsfig}
\usepackage{amsmath}
\usepackage{amssymb}
\usepackage{graphics,epstopdf}
\usepackage{amsfonts}
\usepackage{amsthm}
\usepackage[usenames]{color}

\definecolor{AV}{rgb}{0.65,0.0,0}
\definecolor{GC}{rgb}{0,0.0,0.65}
\definecolor{WS}{rgb}{0,0.65,0}

\usepackage[
      colorlinks=true,
      linkcolor=blue,
      urlcolor=blue,
      filecolor=blue,
      citecolor=blue,
      pdfstartview=FitV,
      pdftitle={},
      pdfauthor={},
      pdfsubject={},
      pdfkeywords={},
      pdfpagemode=None,
      bookmarksopen=true
]{hyperref}

\newcommand{\bm}{\begin{multiline}}
\newcommand{\beq}{\begin{equation}}
\newcommand{\eeq}{\end{equation}}
\newcommand{\beqs}{\begin{eqnarray}}
\newcommand{\eeqs}{\end{eqnarray}}

\newcommand{\ra}{\rightarrow}

\begin{document}

\thispagestyle{empty}

\hfill{}

\hfill{}

\hfill{}

\vspace{32pt}

\begin{center}

\textbf{\Large Accelerating charged and rotating black holes in scalar multipolar universes }

\vspace{48pt}

\textbf{ Cristian Stelea,}\footnote{Corresponding author e-mail: \texttt{cristian.stelea@uaic.ro}}
\textbf{Marina-Aura Dariescu,}\footnote{E-mail: \texttt{marina@uaic.ro}}
\textbf{Vitalie Lungu, }\footnote{E-mail: \texttt{vitalie.lungu@student.uaic.ro}}

\vspace*{0.2cm}

\textit{$^1$ Department of Exact and Natural Sciences, Institute of Interdisciplinary Research,}\\[0pt]
\textit{``Alexandru Ioan Cuza" University of Iasi}\\[0pt]
\textit{11 Bd. Carol I, Iasi, 700506, Romania}\\[.5em]

\textit{$^{2,3}$ Faculty of Physics, ``Alexandru Ioan Cuza" University of Iasi}\\[0pt]
\textit{11 Bd. Carol I, Iasi, 700506, Romania}\\[.5em]

\end{center}

\vspace{30pt}

\begin{abstract}
Recently, Cardoso and Nat\'ario \cite{Cardoso:2024yrb} constructed an exact solution of Einstein-scalar field equations that describes a scalar counterpart of the Schwarzschild-Melvin Universe. In fact, this solution belongs to a more general class of solutions  described by Herdeiro in \cite{Herdeiro:2024oxn}. In this work we show how to further generalize these solutions in presence of acceleration, rotation and various charges. More specifically, we describe general accelerating charged rotating black holes with NUT charge in scalar multipolar universes and present some of their properties.

\end{abstract}

\vspace{32pt}

\setcounter{footnote}{0}

\newpage

\section{Introduction}

Scalar fields  belong to the simplest case of matter fields and they have been consistently used in various fields of modern physics, ranging from General Relativity (GR) to high energy particle physics. For example, in cosmology, scalar fields have been used to model dark energy and dark matter \cite{Quiros:2019ktw} (see also \cite{Faraoni:2004pi}). In the high energy particle physics context the first observational evidence of a scalar particle was obtained at the Large Hadron Collider, in the form of the long sought Higgs boson \cite{ATLAS:2012yve}, \cite{CMS:2012qbp}. Scalar fields also occur naturally in theories involving Kaluza-Klein dimensional reductions, such as string theory. Moreover, in the context of black hole physics in GR the role of scalar fields and their consequences on the no-scalar-hair theorems have been addressed in \cite{Herdeiro:2015waa}. 

Quite recently, in \cite{Cardoso:2024yrb} the authors found a new solution of the Einstein-scalar field equations that describes a black hole immersed into a scalar Universe for which the scalar field is asymptotically aligned with the $z$-axis. This solution has been called the scalar Schwarzschild-Melvin solution, by analogy with the well-known Schwarzschild-Melvin solution in the Einstein-Maxwell theory, for which the magnetic field is aligned asymptotically with the z-axis. 
Later, in \cite{Herdeiro:2024oxn} Herdeiro showed that the scalar Schwarzschild-Melvin solution is a particular case of a class of solutions of the Einstein-scalar field equations that describes black holes in multipolar scalar Universes. This class of solutions belongs to the well-known Weyl-Papapetrou metrics. The background on which the black holes reside is determined entirely by a massless and minimally coupled scalar field. Quite generically, black hole solutions on this scalar background contain naked curvature singularities that reside either on the black hole horizon or in the asymptotic regions. This is to be expected since for a massless and minimally coupled scalar field there are no asymptotically flat regular solutions with scalar fields \cite{Herdeiro:2015waa}.  However, in this class of solutions there do exist particular profiles of the scalar field that correspond to scalar black hole solutions which are regular on the event horizon and in the near horizon region, which means that the naked curvature singularities are in the asymptotic regions. These black hole solutions could still be physically interesting if one considers an appropriate cutoff to avoid the asymptotic regions.

The purpose of this work is to further generalize the Herdeiro solutions in presence of acceleration,  rotation and also electromagnetic and NUT charges. More precisely, in the next section one shows how to simply obtain such a solution in presence of a scalar field. Our approach is based on the symmetries of the dimensionally reduced theory down to three dimensions. Rotating black holes of this kind could also be constructed using the methods described in \cite{Astorino:2014mda}, however, in our approach one could easily include the effect of electric and magnetic charges as well. One should note that the subject of exact solutions of Einstein-scalar field equations in four and higher dimensions has a long history: in fact, in section $2$ we will recover a previously known result discovered first in \cite{Eris:1976xj}. Based on this we were able to easily construct the most general charged and rotating black holes with minimally coupled and massless scalar fields, a solution encompassing various other  accelerating, charged and/or rotating solutions known in literature \cite{Bogush:2020lkp}, \cite{Barrientos:2025abs}, \cite{Jafarzade:2025zbg}, \cite{Kachi:2025fbv}. One should however note that our approach is based on the use of Weyl-Papapetrou coordinates and as such it can only describe black holes with scalar fields up to five dimensions. There are however known methods to add scalar fields to static geometries in higher than five dimensions (see for instance \cite{Maeda:2019tqs} and references therein). Presumably, this kind of solutions could be generalized to higher than five dimensions using the methods of \cite{Barrientos:2025abs}.

The structure of this paper is as follows: in the next section one describes the solution generating technique. In section $3$ one presents the scalar generalization of the Type D accelerating Kerr-Newman black hole with NUT charge in scalar multipolar Universes. We focus on a special solution describing the accelerating and rotating generalization of the Cardoso-Nav\'aro solution and present some of its properties. The final section is dedicated to conclusions and avenues for further work.

\section{The solution generating technique}

As it is well-known by now, Einstein's field equations form a highly nonlinear system of partial differential equations. Solving them in full generality is a daunting task and in order to find new physically meaningful exact solutions one has to rely on various solution generating techniques. Usually, these solutions generating techniques are based on some kind of simplifying  symmetries imposed in the ansatz. Of particular importance is the so-called Weyl-Papapetrou ansatz for the geometry, which is considered to be static and axisymmetric:
\beqs
ds^2&=&-e^{2\psi}dt^2+e^{-2\psi}\bigg[e^{2\gamma}(d\rho^2+dz^2)+\rho^2d\varphi^2\bigg],
\label{wp}
\eeqs
where the functions $\psi$ and $\gamma$ depend on the coordinates $\rho$ and $z$. Performing now a dimensional reduction on the time direction one arrives at the three-geometry of the kind:
\beqs
ds_{3}^2&=&e^{2\gamma}(d\rho^2+dz^2)+\rho^2d\varphi^2.
\label{m3i}
\eeqs
The interesting thing about this geometry is that its Ricci tensor is linear in the function $\gamma$ \cite{Chng:2006gh}, since:
\beqs
R_{\rho\rho}[\gamma]&=&-\left(\partial_z^2\gamma+\partial^2_{\rho}\gamma\right)+\frac{1}{\rho}\partial_{\rho}\gamma,\nonumber\\
R_{\rho z}[\gamma]&=&\frac{1}{\rho}\partial_z\gamma,\nonumber\\
R_{zz}[\gamma]&=&-\left(\partial_z^2\gamma+\partial^2_{\rho}\gamma\right)-\frac{1}{\rho}\partial_{\rho}\gamma.
\label{Ricci3}
\eeqs
One can see now that if one writes down the dimensionally reduced Einstein equations in the form $R_{ij}=T_{ij}-Tg_{ij}$, then for each source $T_{ij}-Tg_{ij}$ one can associate a function $\gamma$ in three dimensions, as long as the matter Lagrangeans of each source are decoupled.

To better understand this situation, let us consider first a static and axisymmetric vacuum solution of Einstein equations as in (\ref{wp}). After performing a dimensional reduction along the time direction one obtains the effective Lagrangian\footnote{Here $g_3=e^{4\gamma}\rho^2$ is the metric determinant of the three geometry, which has Euclidean signature.}:
\beqs
{\cal L}_3&=&\sqrt{g_3}\big[R-2(\partial_{i}\psi)(\partial^{i}\psi)\big].
\label{L3i}
\eeqs
Then Einstein's field equations can be written as $R_{ij}[\gamma]=2(\partial_{i}\psi)(\partial_{j}\psi)$ and they lead to the following equations for the $\gamma$ function:
\beqs
\partial_{\rho}\gamma&=&\rho\big[(\partial_{\rho}\psi)^2-(\partial_{z}\psi)^2\big],\nonumber\\
\partial_{z}\gamma&=&2\rho(\partial_{\rho}\psi)(\partial_{z}\psi).
\label{gamma3eq}
\eeqs
On the other hand, the equation of motion for the scalar field $\psi$ as derived from the Lagrangean (\ref{L3i}) leads to:
\beqs
\Delta\psi&=&e^{2\gamma}\bigg[\partial^2_{\rho}\psi+\partial^2_{z}\psi+\frac{1}{\rho}\partial_{\rho}\psi\bigg]=0.
\eeqs
Since $e^{2\gamma}\neq 0$ this means that the scalar $\psi$ can be considered as a solution of the three-dimensional Laplace equation on an Euclidean geometry corresponding to $\gamma=0$ in (\ref{m3i}) \cite{Chng:2006gh}. Therefore Einstein’s equations in four dimensions for a static axisymmetric background are now essentially reduced to finding a solution of Laplace’s equation on flat space.

Consider now a particular background of the form:
\beqs
ds^2&=&-dt^2+e^{2\mu}[d\rho^2+dz^2]+\rho^2d\varphi^2,
\eeqs
solution of the Einstein-scalar field equations derived from the Lagrangean:
\beqs
{\cal L}_4&=&\sqrt{-g}\big[R-2(\partial_{\mu}\phi)(\partial^{\mu}\phi)\big].
\label{L4sc}
\eeqs
After performing a dimensional reduction along the time direction one obtains the three-geometry of the form:
\beqs
ds_{3}^2&=&e^{2\mu}(d\rho^2+dz^2)+\rho^2d\varphi^2.
\label{m3s}
\eeqs
with the effective Lagrangian:
\beqs
{\cal L}_3&=&\sqrt{g_3}\big[R-2(\partial_{i}\phi)(\partial^{i}\phi)\big].
\label{L3s}
\eeqs
Then Einstein's field equations can be written as $R_{ij}[\mu]=2(\partial_{i}\phi)(\partial_{j}\phi)$ and they lead to the same functional equations for the $\mu$ function as in (\ref{gamma3eq}):
\beqs
\partial_{\rho}\mu&=&\rho\big[(\partial_{\rho}\phi)^2-(\partial_{z}\phi)^2\big],\nonumber\\
\partial_{z}\mu&=&2\rho(\partial_{\rho}\phi)(\partial_{z}\phi).
\label{mu3eq}
\eeqs
 The equation of motion for the scalar field $\phi$ shows that it is a harmonic function in the  auxiliary flat background (\ref{m3s}) (with $\mu=0$). The general solution for this equation can be expressed as \cite{Herdeiro:2024oxn} -\cite{Astorino:2021boj}:
\beqs
\phi&=&\frac{a_0}{R}+\sum_{l=1}^{\infty}\left(\frac{a_l}{R^{l+1}}+b_lR^l\right)P_l,
\label{scalgen}
\eeqs
 where $R=\sqrt{\rho^2+z^2}$ and $P_l$ is the $l$-Legendre polynomial with argument $\frac{z}{R}$. Here $a_0$, $a_l$ and $b_l$ are constants.
 
 One can also integrate (\ref{mu3eq}) to find the function $\mu$ \cite{Herdeiro:2024oxn} -\cite{Astorino:2021boj}:
 \beqs
 \mu&=&\sum_{l, p=1}^{\infty}\bigg[\frac{(l+1)(p+1)a_la_p}{(l+p+2)R^{l+p+2}}\left(P_{l+1}P_{p+1}-P_lP_p\right)+\frac{lpb_lb_pR^{l+p}}{l+p}\left(P_lP_p-P_{l-1}P_{p-1}\right)\bigg].
 \label{muform}
 \eeqs
 
One can now combine the two solutions as follows: consider a four-dimensional geometry sourced by an Einstein-scalar Lagrangean of the form (\ref{L4sc}). If one assumes the geometry to belong to the Weyl-Papapetrou form:
\beqs
ds^2&=&-e^{2\psi}dt^2+e^{-2\psi}\bigg[e^{2\lambda}(d\rho^2+dz^2)+\rho^2d\varphi^2\bigg],
\label{wpf}
\eeqs
then, after performing a dimensional reduction along the time direction one obtains a three-geometry of the form:
\beqs
ds_{3}^2&=&e^{2\lambda}(d\rho^2+dz^2)+\rho^2d\varphi^2,
\label{m3sc}
\eeqs
sourced by a three-dimensional Lagrangean:
\beqs
{\cal L}_3&=&\sqrt{g_3}\big[R-2(\partial_{i}\psi)(\partial^{i}\psi)-2(\partial_{i}\phi)(\partial^{i}\phi)\big].
\label{L3s}
\eeqs
The Einstein's field equations can now be written as:
\beqs
R_{ij}[\lambda]&=&2(\partial_{i}\psi)(\partial_{j}\psi)+2(\partial_{i}\phi)(\partial_{j}\phi)=R_{ij}[\gamma]+R_{ij}[\mu]
\label{einst3s}
\eeqs
and therefore, since the expression of the Ricci tensor in (\ref{Ricci3}) is linear in the function $\lambda$, one can simply take $\lambda=\gamma+\mu$ to satisfy the Einstein equations in three dimensions. The remaining equations of motion derived from the Lagrangean (\ref{L3s}) show that the functions $\psi$ and $\phi$ are harmonic functions $\Delta\psi=\Delta\phi=0$, while the functions $\gamma$ and $\mu$ can be found by integrating (\ref{gamma3eq}) and (\ref{mu3eq}) (which are now equivalent to Einstein's equations (\ref{einst3s})).  In conclusion, for every static and axisymmetric vacuum solution of the form (\ref{wp}) one can find a scalar modification in the form:
\beqs
ds^2&=&-e^{2\psi}dt^2+e^{-2\psi}\bigg[e^{2\gamma+2\mu}(d\rho^2+dz^2)+\rho^2d\varphi^2\bigg],
\label{wpfinst}
\eeqs
where the scalar field $\phi$ is given by (\ref{scalgen}) and the function $\mu$ is given by (\ref{muform}). This is the class of solutions considered by Herdeiro in \cite{Herdeiro:2024oxn}. In the followings we shall show that it is easy to generalize this construction in the presence of rotation and various other charges.

Consider first a charged and rotating generalization of the Weyl-Papapetrou ansatz of the form:
\beqs
ds^2&=&-e^{2\psi}(dt+{\cal A})^2+e^{-2\psi}\bigg[e^{2\gamma}(d\rho^2+dz^2)+\rho^2d\varphi^2\bigg],
\label{wprot}
\eeqs
where ${\cal A}={\cal A}_{\varphi}d\varphi$. Let us assume that, in four dimensions, this geometry is a solution of the field equations derived from the Lagrangean:
\beqs
{\cal L}_4&=&\sqrt{-g}\big[R-F^2\big],
\label{Lag4d}
\eeqs
where one considers the Maxwell field $F=dA$, with $A=\chi dt+A_{\varphi}d\varphi$. Performing now a dimensional reduction on the time direction one obtains the three-dimensional Lagrangean:
\beqs
{\cal L}_3&=&\sqrt{g_3}\bigg[R_3-2(\partial \psi)^2+e^{4\psi}({\cal F})^2-e^{2\psi}F_{(2)}^2+2e^{-2\psi}(\partial\chi)^2\bigg],
\label{Lag3drot}
\eeqs
while the three-geometry is given by (\ref{m3i}). Here we defined $F_{(2)}=dA_{\varphi}\wedge d\varphi-d\chi\wedge {\cal A}$ and ${\cal F}=d{\cal A}$, while we denoted $(\partial \psi)^2=(\partial_i\psi)(\partial^i\psi)$, ${\cal F}^2={\cal F}_{ij}{\cal F}^{ij}$, $F_{(2)}^2=F_{(2)ij} F_{(2)}^{ij}$ and $(\partial \chi)^2=(\partial_i\chi)(\partial^i\chi)$. One can write the three-dimensional Einstein equations in the following form:
\beqs
R_{ij}[\gamma]&=&T_{ij}^{(1)}=2(\partial_{i}\psi)(\partial_{j}\psi)-2e^{4\psi}\bigg[{\cal F}_{ik}{\cal F}_j^{~k}-\frac{1}{2}g_{ij}{\cal F}_{kl}{\cal F}^{kl}\bigg]-2e^{-2\psi}(\partial_{i}\chi)(\partial_{j}\chi)+\nonumber\\
&+&2e^{2\psi}\bigg[F_{(2)ik}F_{(2)j}^{~k}-\frac{1}{2}g_{ij}F_{(2)kl} F_{(2)}^{kl}\bigg].
\label{Ein3rot}
\eeqs
The remaining field equations in three dimensions are the field equations for the matter fields, as derived from the Lagrangean (\ref{Lag3drot}):
\beqs
\nabla_i\left(e^{4\psi}{\cal F}^{ij}\right)&=&0,~~~~~\nabla_i\left(e^{2\psi}F_{(2)}^{ij}\right)=0,~~~~~
\nabla_i\left(e^{-2\psi}\nabla^i\chi\right)=0,\nonumber\\
\Delta\psi&=&\frac{1}{2}e^{2\psi}F_{(2)}^2-e^{4\psi}{\cal F}^2-e^{-2\psi}(\partial\chi)^2.
\label{eomatter}
\eeqs
Together, the system of equations given in (\ref{Ein3rot}) and (\ref{eomatter}) is equivalent to the system of equations derived from the Einstein-Maxwell Lagrangean (\ref{Lag4d}).

Consider now a four-dimensional geometry that should be a solution of the Einstein-Maxwell-scalar system described by the Lagrangean:
\beqs
{\cal L}_4&=&\sqrt{-g}\big[R-2(\partial_{\mu}\phi)(\partial^{\mu}\phi)-F^2\big].
\label{Lag4dfull}
\eeqs
In general, let us assume that one can write it in the Weyl-Papapetrou form:
\beqs
ds^2&=&-e^{2\psi}(dt+{\cal A})^2+e^{-2\psi}\bigg[e^{2\lambda}(d\rho^2+dz^2)+\rho^2d\varphi^2\bigg].
\label{wpfull}
\eeqs
Performing now a dimensional reduction on the timelike direction one is lead to the three-dimensional geometry of the form (\ref{m3sc}), which should be a solution of the field equations derived from the dimensionally reduced Lagrangean:
\beqs
{\cal L}_3&=&\sqrt{g_3}\bigg[R_3-2(\partial \psi)^2+e^{4\psi}({\cal F})^2-e^{2\psi}F_{(2)}^2+2e^{-2\psi}(\partial\chi)^2-2(\partial_{i}\phi)(\partial^{i}\phi)\bigg].
\label{Lag3full}
\eeqs
In this case Einstein's field equations can be cast in the form:
\beqs
R_{ij}[\lambda]&=&T^{(1)}_{ij}+T^{(2)}_{ij},
\eeqs
where $T^{(1)}_{ij}=R_{ij}[\gamma]$ is listed in (\ref{Ein3rot}), while $T^{(2)}_{ij}=2(\partial_{i}\phi)(\partial_{j}\phi)=R_{ij}[\mu]$. Thence one obtains $R_{ij}[\lambda]=R_{ij}[\gamma]+R_{ij}[\mu]$ and therefore one can simply take $\lambda=\gamma+\mu$ to simply satisfy Einstein's field equations in three dimensions.

In order to check that the remaining field equations for the various matter fields in (\ref{Lag3full}) are satisfied, one should note that the scalar Lagrangean term for the scalar field $\phi$ is decoupled from the rest of the matter fields and it only adds the field equation $\Delta\phi=0$ to them, while all the remaining field equations are the same as in (\ref{eomatter}) and they are satisfied as well if the initial solution (\ref{wprot}) satisfies them. 

In conclusion, starting with a known stationary solution (\ref{wprot}) of the Einstein-Maxwell equations then the new solution of the field equations derived from the Lagrangean (\ref{Lag4dfull}) can be written in the form:
\beqs
ds^2&=&-e^{2\psi}(dt+{\cal A})^2+e^{-2\psi}\bigg[e^{2\gamma+2\mu}(d\rho^2+dz^2)+\rho^2d\varphi^2\bigg],
\label{wpfull4d}
\eeqs
where the scalar field $\phi$ takes the general form (\ref{scalgen}), while the function $\mu$ is given by (\ref{muform}). A similar theorem was previously proven by other means in \cite{Eris:1976xj}.

\section{The accelerating Kerr-Newman-NUT solution with a multipolar scalar field}

As an example of the solution generating technique described in the previous section, let us consider the newly discovered Type D accelerating Kerr-Newman-NUT solution in four dimensions (see for instance \cite{Astorino:2024bfl}, \cite{Ovcharenko:2024yyu})\footnote{Here we have chosen the value of the constant $C_f=\frac{1}{(1+\alpha^2 a^2)}$.}:
\beqs
ds^2&=&\frac{1}{\Omega^2}\bigg[-\frac{\Delta_r}{\Sigma}\left(A dt-B d\varphi\right)^2+\frac{\Sigma}{(1+\alpha^2a^2)}\left(\frac{dr^2}{\Delta_r}+\frac{dx}{\Delta_x}\right)^2+\nonumber\\
&+&\frac{\Delta_x}{\Sigma}\left(Cdt+D d\varphi\right)^2\bigg],
\label{accKN}
\eeqs
where we defined:
\beqs
\Omega&=&1-\alpha r x, ~~~A=1+\alpha^2(l^2-a^2)x^2, ~~~B=a+2lx+ax^2,\nonumber\\
C&=&a+2\alpha l r+\alpha^2 a r^2,~~~D=(l^2-a^2)+r^2\nonumber\\
\Delta_r&=&(1-\alpha^2 r^2)\big[(r-m)^2-\sigma^2\big],~~~\Delta_x=(1-x^2)\big[(1-\alpha m x)^2-\alpha^2x^2\sigma^2\big],\nonumber\\
\Sigma&=&(l+ax)^2+2\alpha l(a+2lx+ax^2)r+\alpha^2(a^2-l^2)^2x^2+\big[1+\alpha^2(a+lx)^2\big]r^2,\nonumber\\
\sigma&=&\sqrt{m^2+l^2-a^2-e^2-g^2}.
\eeqs
Here $\alpha$ is the acceleration parameter, $m$ is the mass parameter, $a$ is the rotation parameter, $l$ is the NUT charge, while $e$ and $g$ are the electric and magnetic charges. The components of the electromagnetic Maxwell field $A=\chi dt+A_{\varphi}d\varphi$ are listed in \cite{Astorino:2024bfl}, \cite{Ovcharenko:2024yyu}, \cite{Ovcharenko:2025fxg} and, for simplicity, we will not list them here since the addition of a scalar field in this background has no effect on the Maxwell field components. In this geometry one makes use of the radial coordinate $r$ and the coordinate $x$ that can be related to the usual polar angle $\theta$ by using the relation $x=\cos\theta$. If the acceleration parameter is set equal to zero, $\alpha=0$, one recovers the Kerr-Newman-NUT black hole solution. 

As it is well-known, the Kerr-Newman-NUT metric element can be brought into the Weyl-Papapetrou form (\ref{wprot}) by using the following coordinate transformations:
\beqs
\rho&=&\sqrt{\Delta_r}\sin\theta,~~~~~z=(r-m)\cos\theta,
\label{coordinates}
\eeqs
such that:
\beqs
d\rho^2+dz^2&=&(\Delta_r\cos^2\theta+(r-m)^2\sin^2\theta)\left(\frac{dr^2}{\Delta_r}+d\theta^2\right),
\eeqs
while:
\beqs
e^{2\gamma}&=&\frac{\Delta_r-a^2\sin^2\theta}{\Delta_r\cos^2\theta+(r-m)^2\sin^2\theta}.
\eeqs
Then the general Kerr-Newman-NUT solution in a multipolar scalar universe will be given by:
\beqs
ds^2&=&-\frac{\Delta_r}{\Sigma}\bigg[dt-\left(a\cos^2\theta+2l\cos\theta+a)\right)d\varphi\bigg]^2+\Sigma e^{2\mu}\left(\frac{dr^2}{\Delta_r}+d\theta\right)^2+\nonumber\\
&+&\frac{\sin^2\theta}{\Sigma}\big[adt-\left(r^2+(l^2-a^2)\right)d\varphi\big]^2,
\label{gensol}
\eeqs
where the Maxwell field components are listed in \cite{Astorino:2024bfl}, \cite{Ovcharenko:2024yyu}, \cite{Ovcharenko:2025fxg}, the scalar field $\phi$ has the general expression (\ref{scalgen}), while the function $\mu$ is given by (\ref{muform}), using (\ref{coordinates}). Harmonic scalars with growing multipoles (described by terms proportional to $b_l$ in the multipolar expansion (\ref{scalgen})) lead to naked singularities at spatial infinity, while harmonic scalars with decaying multipoles (described by the terms proportional to $a_l$ in (\ref{scalgen})) lead to asymptotic flat geometries with singular horizons, in accordance to the no-hair theory.

Note that the horizon location in our new solution is again determined by the roots of $\Delta_r=0$, while the ergosphere has the same properties as in the Kerr solution. The Ricci scalar for the geometry (\ref{gensol}) is given by $R=2(\partial_{\mu}\phi)(\partial^{\mu}\phi)$ and, therefore, the curvature singularities in the metric are directly related to the divergencies of the scalar field. Moreover, the effects of the scalar field in the metric are encoded using the function $e^{2\mu}$ and it will generically lead to distortions of the black hole horizon. For simplicity, let us set the NUT charge to zero $l=0$. Then the metric induced on the black hole horizon $r=r_H$ is:
\beqs
ds_H^2&=&\Sigma_H e^{2\mu_H}d\theta^2+\frac{\Psi_H\sin^2\theta}{\Sigma_H}d\varphi^2,
\eeqs
where we defined $\Psi_H=(r_H^2+a^2)^2$, $\Sigma_H=r_H^2+a^2\cos^2\theta$, while $\mu_H$ is the value of the function $\mu$ on the black hole horizon. Then the area of the black hole horizon can be written as:
\beqs
Area&=&2\pi(r_H^2+a^2)\int_0^{\pi}e^{\mu_H}\sin\theta d\theta.
\eeqs

Let us now consider the presence of the acceleration parameter $\alpha$. In this case it can be shown that the general Type D accelerating Kerr-Newman-NUT solution (\ref{accKN}) can still be brought into the Weyl-Papapetrou form by employing the following coordinate transformations:
\beqs
\rho&=&\frac{\sqrt{\Delta_r\Delta_x}}{(1+\alpha r x)^2}, ~~~ z=\frac{(\alpha r+x)\big[(r-m)(1+\alpha m x)+\alpha x\sigma^2\big]}{(1+\alpha r x)^2}.
\label{coordweyl}
\eeqs
Therefore, the general Type D accelerating Kerr-Newman-NUT solution in a scalar multipolar universe can be written in the following symmetric form:
\beqs
ds^2&=&\frac{1}{\Omega^2}\bigg[-\frac{\Delta_r}{\Sigma}\left(A dt-B d\varphi\right)^2+\frac{\Sigma}{(1+\alpha^2a^2)}e^{2\mu}\left(\frac{dr^2}{\Delta_r}+\frac{dx}{\Delta_x}\right)^2+\nonumber\\
&+&\frac{\Delta_x}{\Sigma}\left(Cdt+D d\varphi\right)^2\bigg],
\label{accKNscalar}
\eeqs
where once again the scalar field $\phi$ is given in (\ref{scalgen}), while the function $\mu$ is given by (\ref{muform}), using now the coordinate transformations given in (\ref{coordweyl}). The horizons are again determined by the roots of the equation $\Delta_r=0$. One obtains the inner horizon, located at $r_-=m-\sqrt{\sigma}$, the black hole horizon at $r_+=m+\sqrt{\sigma}$, while acceleration horizon is located at $r_{\alpha}=\frac{1}{\alpha}$.

As a particular example, let us consider the scalar profile used in \cite{Cardoso:2024yrb} to construct the scalar-Schwarzschild-Melvin solution. In this case:
\beqs
\phi&=&kz=k\frac{(\alpha r+x)\big[(r-m)(1+\alpha m x)+\alpha x\sigma^2\big]}{(1+\alpha r x)^2}, ~~~~~e^{2\mu}=e^{-k^2\rho^2}=e^{-\frac{k^2\Delta_r\Delta_x}{(1+\alpha r x)^4}}.
\eeqs
If one sets to zero all the charges  and the acceleration parameter $\alpha=0$ one obtains $\sigma^2=m^2-a^2$, which corresponds to the rotating version of the scalar-Schwarzschild-Melvin \cite{Cardoso:2024yrb} in the form previously found in \cite{Sen:1998ftn}:
\beqs
ds^2&=&-\frac{\Delta_r}{\Sigma}\bigg[dt-a\sin^2\theta d\varphi\bigg]^2+\Sigma e^{2\mu}\left(\frac{dr^2}{\Delta_r}+d\theta\right)^2+\frac{\sin^2\theta}{\Sigma}\big[adt-\left(r^2+a^2\right)d\varphi\big]^2,
\label{gensolKerr}
\eeqs
where $\Delta_r=r^2-2mr+a^2$ and $\Sigma=r^2+a^2\cos^2\theta$. It is easy to check that it satisfies the Einstein-scalar equations $R_{\mu\nu}=2(\partial_{\mu}\phi)(\partial_{\nu}\phi)$.  If $a=0$ it further reduces to the solution in \cite{Cardoso:2024yrb}.
Its Ricci scalar can be simply expressed as:
\beqs
R&=&\frac{2k^2e^{k^2\Delta_r\sin^2\theta}\left(\Delta_r+(m^2-a^2)\sin^2\theta\right)}{r^2+a^2\cos^2\theta}.
\eeqs
Besides the usual curvature singularities presented in the Kerr metric one can see that the backreaction of the scalar field leads to another curvature singularity located in the asymptotic region as $r\ra\infty$. Note that in this case the area of the black hole horizon is the same as the area computed for the Kerr black hole since $\mu_H=0$ on the black hole horizon.

\section{Conclusions}

Recently, Cardoso and Nat\'ario \cite{Cardoso:2024yrb} constructed an exact solution of Einstein-scalar field equations that describes a scalar counterpart of the Schwarzschild-Melvin Universe. In fact, this solution belongs to a more general class of solutions  described by Herdeiro in \cite{Herdeiro:2024oxn}. These solutions describe black holes in the so-called scalar multipolar Universes, which are determined in terms of the multipolar terms appearing in the scalar field expansion (\ref{scalgen}). Generically, these black hole solutions present either singularities on the black hole horizon, or singularities in the asymptotic regions, rendering them non-asymptotically flat. However, for special profiles of the scalar fields the singularities can be pushed out to the asymptotic region, keeping the black hole horizon and the regions near it nonsingular. 

 In our work we further generalized this class of exact solutions of the Einstein-scalar system to include acceleration, rotation and various other charges. More specifically, in section $2$ we presented our solution generating technique that allowed us to combine any stationary exact solution of the Einstein-Maxwell system with axial symmetry with the backreaction of a minimally coupled scalar field. Our technique is based on the symmetries of the dimensionally reduced Lagrangeans down to three dimensions along the timelike direction and it relies on the properties of the Weyl-Papapetrou ansatz for the metric. Our solution generating technique can be regarded as an alternative way of deriving  a result previously known in the literature \cite{Eris:1976xj}. However, we do believe that our technique can be easily extended to work in context of general sigma-models, as arising in the dimensional compactification of string theories.

As an example we derived the general accelerating Type D Kerr-Newman-NUT solution in the scalar multipolar universes. In absence of the multipolar scalar field this solution was recently derived in \cite{Astorino:2024bfl} (see also \cite{Ovcharenko:2024yyu}, \cite{Ovcharenko:2025fxg}). Setting all the charges to zero and for a specific profile of the scalar field we showed how one can recover the accelerating and rotating version of the Cardoso and Nat\'ario solution \cite{Cardoso:2024yrb}. If the acceleration parameter is set to zero, $\alpha=0$ one recovers a rotating solution previously derived in  \cite{Sen:1998ftn}.

One interesting application of our technique is to use the effect of the scalar field to balance multi-black hole systems. In absence of the scalar field such systems are not in equilibrium and they generically present various conical defects in order to accommodate the presence of other forces needed to balance the attractive gravity effects.  For example, in four and higher dimensions the equilibrium conditions in presence of Maxwell fields have been considered in \cite{Chng:2008sr} - \cite{Stelea:2017pdk} (and references therein). These  multi-black hole systems are usually constructed in Weyl-Papapetrou coordinates and therefore they can be used as seeds in our solution generating technique. By adding carefully chosen scalar profiles one can achieve equilibrium as in \cite{Vigano:2022hrg}, \cite{Astorino:2021boj}. The use of scalar fields to locally balance systems of multi-black holes is not new, as they have been used for instance in \cite{Herdeiro:2023roz}, \cite{Herdeiro:2023mpt} in order to achieve this goal.

Another straightforward generalization of our results is to obtain the corresponding solutions in five dimensions, such as the Kerr-Myers solution with scalar fields, or to study the effect of the scalar fields in balancing the static black rings in five dimensions. One first step in this direction was taken in \cite{Barrientos:2025abs}, where the five dimensional Myers-Perry black hole with a scalar field was first constructed.

Note also that our solution generating technique can be easily extended in presence of other minimally coupled scalar fields (or more generally to a scalar sigma model in four dimensions).

Work on these matters is in progress and it will be presented elsewhere.

\vspace{10pt}


\end{document}